\documentclass[superscriptaddress,aps,amsmath,amssymb,twocolumn,prb]{revtex4-1}

\usepackage{graphicx}
\usepackage{bm}
\usepackage{cases}
\usepackage{amsmath}
\usepackage{color}
\usepackage{soul}

\newcommand{\du}{d_{\uparrow}}
\newcommand{\dd}{d_{\downarrow}}

\newcommand{\red}[1]{\textcolor{black}{#1}}


\begin{document}
\title{Spin backflow: a non-Markovian effect on spin pumping}

\author{Kazunari Hashimoto}
\affiliation{
Graduate School of Interdisciplinary Research, University of Yamanashi, Kofu 400-8511, Japan
}
\author{Gen Tatara}
\affiliation{
RIKEN Center for Emerging Matter Science (CEMS), 2-1 Hirosawa, Wako 351-0198, Japan
}
\author{Chikako Uchiyama}
\affiliation{
Graduate School of Interdisciplinary Research, University of Yamanashi, Kofu 400-8511, Japan
}
\affiliation{
National Institute of Informatics, 2-1-2 Hitotsubashi, Chiyoda-ku, Tokyo 101-8430, Japan
}

\date{\today}

\begin{abstract}
The miniaturization of spintronic devices, specifically, nanoscale devices employing spintronics, has attracted intensive attention from a scientific as well as engineering perspective.
In this paper, we study a non-Markovian effect on spin pumping to describe spin current generation driven by the magnetization of arbitrary precession frequency in a quantum dot attached to an electron lead.
Although the Markovian approximation can be used when driving is sufficiently slow compared with relaxation times in electron tunneling, recent developments in nano-spintronic devices show that we need to include non-Markovian effects.
In contrast to the one-way-only nature of the spin current generation under Markovian dynamics, we find that non-Markovian dynamics exhibit a temporal backflow of spin, called spin backflow for brevity.
We capture the phenomenon by introducing its quantifier, and show that the backflow reduces the amount of spin current significantly when the frequency exceeds the relaxation rate.
This prevents an unphysical divergence of the spin current in the high frequency limit that occurs under the Markovian approximation.
We believe our analysis provides an understanding of the spin pumping particularly in regard to producing a more efficient spin current generation over shorter time scales by going beyond the conventional Markovian approximation.
\end{abstract}

\maketitle

\section{Introduction}
Controlling the electron transport in nano-systems represents a promising advance for future electronics.
Its major application is the single-electron transistor \cite{fulton87}, which would enable extreme downsizing and ultra-low-power consumption of computing devices.
An ambitious research field with this direction in mind seeks to incorporate magnetic components into nano-electronic devices \cite{dempsey11,awshalom13}.
It aims to boost conventional nano-electronics devices by exploiting the spin degrees of freedom in addition to the electronic charge \cite{chen02,yang08,bogani08}.

The generation of spin current is the important aspect in nano-spintronics.
To date, numerous efforts have been made to realize spin pumping in nano-systems \cite{mucciolo02,wang03,zhang03,cota05,splettstoesser08,braun08,hattori08,riwar10,fransson10,winkler13,rojekpssb13,rojekprb13,jahn13,chen15,nakajima15,tatara16}.
\red{A typical protocol uses magnetization precession\cite{tserkovnyakprl02,tserkovnyakprb02}, which has been implemented in bulk systems, consists of a ferromagnet attached to a normal metal \cite{maekawa} as well as superconducting materials \cite{jeon18,jeon19}}.
Because of its wide range of application, it keeps attracting growing interests from both theoretical and experimental points of view.
In contributing to these attempts, we have focused on a minimum model describing spin pumping in a nano-system consisting of an electron lead attached to a two-level system (quantum dot) subjected to a rotating magnetic field \cite{wang03,hattori08,fransson10,tatara16}.

In conventional studies on the minimum model, spin pumping has been formulated using the adiabatic approximation, which requires the rotation frequency of the magnetic field $\Omega$ to be small compared with the characteristic energy scale $\delta E$ over which the stationary scattering property of an electron by the quantum dot changes significantly, i.e., $\Omega\ll\delta E/\hbar$ \cite{moskalets}.
Underlying this condition is an implicit assumption, specifically, the relaxation time $\tau_r$ of the electron distribution in the dot by tunneling to the lead is infinitely slow compared with the rotation, $\tau_r^{-1}\ll\Omega$.
Because setting the relaxation time to infinity is impossible, we studied the effect of its finiteness in Ref.~\cite{hashimoto17} by evaluating the non-adiabatic effect up to $\Omega\lesssim\tau_r^{-1}$ formulated subject to the Born--Markov approximation; see for example\cite{breuer}.
In consequence, we showed that spin pumping is an entirely non-adiabatic effect.
We also found that the non-adiabatic spin current depends linearly on $\Omega$ in a low-frequency regime \cite{tatara_note} and exhibits an oscillatory dependence on $\Omega$, indicating an enhancement of the spin current. 

Despite the treatment in Ref.~\cite{hashimoto17} describing spin pumping with finite precession frequency, its range of applicability is limited to a relatively slow precession because of the Markovian approximation.
The approximation is only valid when the time scale of the relevant dynamics is sufficiently longer than the relaxation time of the dot as well as the correlation time of the lead \cite{breuer}.
Therefore, breakdown occurs for a rapid precession when the relaxation time is exceeded, which often occurs for nano-spintronics systems.
Indeed, in a single molecule magnet system, the rotation frequency of its magnetic core ($\nu\approx10$~GHz) exceeds the relaxation rate ($\gamma_r\approx1\sim10$ ${\rm s}^{-1}$) \cite{note}.
In the present paper, we examine the non-Markovian effect on spin pumping by removing the Markovian approximation from its formulation.

Among several non-Markovian effects \cite{wolf08,breuer09,breuer12,ravis10,lu10,luo12,lorenzo13,bylicka14,chruscinski14,ravis14,breuer16,guarnieri16}, we focus on those revealed as backflow \cite{breuer16,guarnieri16}.
\red{Backflow reflects a partially reversible dynamics of an open system within a time interval in which the memory of the initial condition remains and the dynamics is coherent.}
It allows a back-and-forth transfer of physical quantities such as information \cite{breuer16} and energy \cite{guarnieri16} unlike the one-way-only transfer under Markovian dynamics.
Although conventional studies on the backflow treat undriven systems, it may significantly affect electron transport in a constantly driven system because non-Markovian effects dominate the initial stage of the relaxation process following a given external disturbance.
Now the question arises: what is the role of backflow in a constantly driven system such as in spin pumping?
\red{To answer this question, we formulate the spin pumping by using the full counting statistics, which enables us to describe electron transfer dynamics during the time interval between two successive measurements of electron number \cite{guarnieri16,esposito,uchiyama14}.
By including the non-Markovian effect to the dynamics, we obtain a short-time behavior description of partial reversibility allowing spin transfer back from lead to dot, which we call spin backflow.}
We find that the non-Markovian dynamics enables a physically reasonable description of spin pumping over the whole frequency range.

\section{Model}
\begin{figure}[t]
\centering
\includegraphics[keepaspectratio, scale=0.38, angle=0]{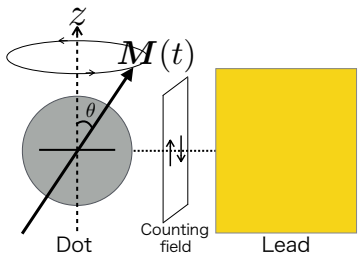}
\vspace{-15pt}
\caption{Schematic drawing of the minimum model. The model consists of a ferromagnetic quantum dot attached to an electron lead. The dot has a dynamic magnetization ${\bm M}(t)$ that rotates around the $z$-axis with a period ${\cal T}$. The number of transferred electrons with spin magnetic moment $\uparrow$ ($\downarrow$) is captured by the counting field (see {\it Formalism}).}
\vspace{-17pt}
\end{figure}
We consider a minimum model of spin pumping (Fig.~1) that describes a quantum dot with a dynamic magnetization attached to an electron lead \cite{tatara16,hashimoto17}.
In the quantum dot, the electron is spin polarized because of the {\it s-d} exchange interaction with the magnetization and is represented by a two-component creation and annihilation operators ${\bm d}^{\dagger}=(\du^{\dagger},\dd^{\dagger})$ and ${\bm d}$, where $\uparrow$ or $\downarrow$ represents the direction of the spin magnetic moment of the electron parallel or antiparallel to the $z$-axis.

The Hamiltonian $H(t)=H_{{\rm d}}(t)+H_{{\rm l}}+H_{{\rm t}}$ contains three terms: $H_{{\rm d}}(t)$ describing the dot is defined by $H_{{\rm d}}(t)={\bm d}^{\dagger}[\epsilon_{{\rm d}}-{\bm M}(t)\cdot{\bm\sigma}]{\bm d}$, where $\epsilon_{{\rm d}}$ is the unpolarized energy of a dot electron, ${\bm M}(t)\equiv M(\sin\theta\cos\phi(t),\sin\theta\sin\phi(t),\cos\theta)$, and ${\bm\sigma}=(\sigma_x,\sigma_y,\sigma_z)$ is the vector of Pauli matrices.
The electron lead is described by the term $H_{{\rm l}}=\sum_{\sigma=\uparrow,\downarrow}\sum_{k}\epsilon_{k}c_{\sigma,k}^{\dagger}c_{\sigma,k}$, where $c_{\sigma,k}^{\dagger}$ and $c_{\sigma,k}$ with $\sigma=\uparrow$ or $\downarrow$ the creation and annihilation operators of a lead electron with energy $\epsilon_{k}$.
The dot--lead interaction is assumed to be spin conserving with $H_{{\rm t}}=\sum_{\sigma}\sum_{k}\hbar v_{k}(d_{\sigma}^{\dagger}c_{\sigma,k} +c_{\sigma,k}^{\dagger}d_{\sigma})$, where $\hbar v_{k}$ is the coupling strength, which we assume to be weak.
\red{
In the following, we apply full counting statistics (FCS)\cite{esposito} to evaluate the number of transferred electrons with spin $\sigma$ through projective measurements of the electron number in the lead represented by $N_{\sigma}\equiv\sum_{k}c_{\sigma,k}^{\dagger} c_{\sigma,k}$.
Defining an outcome of the projective measurement at time $t$ as $n_{\sigma,t}$, we discuss electron dynamics under the spin pumping.}

\section{formalism}
Let us briefly summarize how we apply the FCS to formulate spin pumping.
Details are presented in Supplementary Materials.

\red{The FCS is based on the joint probability of outcomes of two successive projective measurements.
It provides the statistical average of the number of transferred electrons through the unitary time evolution under the dot--lead interaction between measurements over the initial states of the total system.
Using the joint probability, we obtain the probability density of the difference between the two outcomes at time $t_{i}$ and a later time $t_{i+1}(=t_{i}+\delta t)$, which we define as $P(\Delta n_{\sigma,i})$ for the difference $\Delta n_{\sigma,i}(\equiv n_{\sigma,t_{i+1}}-n_{\sigma,t_{i}})$. The sign of $\Delta n_{\sigma,i}$ is chosen to be positive when electrons are transferred from dot to lead.
To obtain cumulants of $\Delta n_{\sigma,i}$, it is convenient to use the generating function, the Fourier transform of $P(\Delta n_{\sigma,i})$, i.e., $G(\lambda_{\sigma})\equiv\int^{\infty}_{-\infty}P(\Delta n_{\sigma,i}) e^{i\lambda_{\sigma,i}\Delta n_{\sigma,i}}d\Delta n_{\sigma,i}$, where the parameter $\lambda_{\sigma}$ is called the counting field.
It gives the first cumulant (mean value) as $\langle\Delta n_{\sigma,i}\rangle=\partial G(\lambda_{\sigma})/\partial(i\lambda_{\sigma})|_{\lambda_{\sigma}}=0$.}

\red{Our next task is to describe the time evolution of $G(\lambda_{\sigma})$.
Assuming that the initial state associated with the joint probability is factorized between dot and lead, and the lead is in a diagonal state with choosing a Gibbs ensemble, we can rewrite $G(\lambda_{\sigma})$ with a traced quantity over the total system where the unitary time evolution operator is modified to include $\lambda_{\sigma}$  (see eq.(S.11) in supplementary material).
Taking the trace procedure in the joint probability for the lead first, we cast the reduced operator for the dot system in the form of a generalized master equation.
In this work, we take the time-convolutionless quantum master equation\cite{tclpapers,uchiyama99} to obtain $\partial\rho^{(\lambda_{\sigma})}(t)/\partial t=\xi^{(\lambda_{\sigma})}(t) \rho^{(\lambda_{\sigma})}(t)$.\cite{uchiyama14,note4}}
\red{The super-operator $\xi^{(\lambda_{\sigma})}(t)$ is expanded as a sum of ``ordered cumulants'' of the interaction Hamiltonian $H_{{\rm t}}$ up to infinite order.}
Taking leading terms up to second-order, we have $\xi^{(\lambda_{\sigma})}(t)\rho =-i\hbar^{-1}[H_{{\rm d}},\rho]+K^{(\lambda_{\sigma})}_{2}(t)\rho$, where $K^{(\lambda_{\sigma})}_{2}(t)\rho=-\hbar^{-2}\int^{t}_{0}d\tau{\rm Tr}_{{\rm l}}[H_{{\rm t}},[H_{{\rm t}}(-\tau),\rho\otimes\rho^{{\rm eq}}_{{\rm l}}]_{\lambda_{\sigma}}]_{\lambda_{\sigma}}$ is the memory kernel with definitions $H_{{\rm t}}(t)\equiv e^{i(H_{{\rm d}}+H_{{\rm l}})t/\hbar}H_{{\rm t}}e^{-i(H_{{\rm d}}+H_{{\rm l}})t/\hbar}$, $[A,B]_{\lambda_{\sigma}}\equiv A^{(\lambda_{\sigma})}B-BA^{(-\lambda_{\sigma})}$, and $A^{(\lambda_{\sigma})}\equiv e^{i\lambda_{\sigma}N_{\sigma}/2}Ae^{-i\lambda_{\sigma}N_{\sigma}/2}$.
The time dependence of the memory kernel reflects the finiteness of the correlation time of the dot--lead interaction, which allows us to describe the non-Markovian dynamics.
Using the generalized master equation, we obtain $\langle\Delta n_{\sigma,i}\rangle =\int^{t_{i+1}}_{t_{i}}J_{\sigma}(s)ds$ with the inertial flow of electrons, $J_{\sigma}(t)\equiv{\rm Tr}_{{\rm d}} [\partial\xi^{(\lambda_{\sigma})}(t)\partial(i\lambda_{\sigma})|_{\lambda_{\sigma}=0}\rho^{(0)}(t)]$ , where ${\rm Tr}_{{\rm d}}$ denotes the trace operation over the states of the dot.

To formulate spin pumping based on the above framework, we consider a step-like change in the direction of ${\bm M}(t)$ around the $z$-axis; specifically, dividing the period ${\cal T}$ into $N$ intervals, $t_{i}\leq t\leq t_{i+1}$ ($i=1,2,\cdots,N$) with $t_{1}=0$ and $t_{N+1}={\cal T}$, fixing the direction of ${\bm M}(t)$ during each interval, and changing $\phi$ at each $t_{i}$ discretely with substitution $\phi_{i}=\phi_{i-1}+\delta\phi$ with $\phi_{0}=0$, $\phi_{N}=2\pi$ and $\delta\phi\equiv2\pi/N$.
Given that the total density matrix is factorized at each $t_{i}$, we obtain the mean number $\langle\Delta n_{\sigma,i}\rangle$.
In the following, we use the \red{instantaneous spin current} defined by
\red{\begin{equation}\label{eq:spinflow}
J_{{\rm spin}}(t)=J_{\uparrow}(t)-J_{\downarrow}(t).
\end{equation}}
Its time integration over one period provides a \red{temporal average of spin current},
\red{\begin{equation}\label{eq:spincurrent}
I_{{\rm spin}}\equiv\frac{1}{{\cal T}}\int^{{\cal T}}_{0}J_{{\rm spin}}(t)dt.
\end{equation}}

\section{Spin backflow}
We introduce the concept of spin backflow, which is different from the spin-current backflow introduced in Ref.~\cite{tserkovnyakprb02}; see {\it Discussions}.
As shown above, the memory kernel, $K^{(\lambda_{\sigma})}_2(t)$, in our formalism includes the finite correlation time of the dot--lead interaction.
\red{The time dependence enables us to describe the time interval in which the memory of the initial condition remains and the electron dynamics is coherent, called partial reversibility in non-Markovian dynamics.}
Partial reversibility, allowing the back-and-forth transfer of an electron, is revealed with the sign reversal of $K^{(\lambda_{\sigma})}_2(t)$, which turns out to be the dynamical change in the direction of the instantaneous spin current $J_{{\rm spin}}(t)$.
We call this return of the electron spin from the lead a {\it spin backflow}.
The time reversible spin exchange has been neglected in the conventional treatment with the Markovian approximation, for which the time-dependence is removed by taking the long-time limit of the memory kernel, specifically, $\lim_{t\to\infty}K^{(\lambda_{\sigma})}_{2}(t)$.
As the approximated memory kernel is time-independent, Markovian dynamics is characterized by the one-way-only transfer of electron spin.

The spin backflow is captured by monitoring the temporal sign change of the instantaneous spin current $J_{{\rm spin}}(t)$, Eq.~(\ref{eq:spinflow}) \cite{guarnieri16}.
When $J_{{\rm spin}}(t)$ is positive, spin is transferred from dot to lead; conversely, when $J_{{\rm spin}}(t)$ is negative, spin is transferred from lead to dot.
In contrast, under the Markovian approximation, we expect that the sign of the $J_{{\rm spin}}(t)$ remains the same during its time evolution.

\section{Numerical results}
Let us now analyze spin backflow by numerically evaluating the instantaneous spin current $J_{{\rm spin}}(t)$ as well as the temporal average of spin current $I_{{\rm spin}}$, Eq.~(\ref{eq:spincurrent}).
In each instance, we also present numerical results obtained subject to the Markovian approximation as a reference for comparison with the non-Markovian analysis.

To describe the dot--lead coupling, we use the Ohmic spectral density with an exponential cutoff $v(\omega)\equiv\sum_{k} v_{k}^{2}\delta(\omega-\omega_{k})=\lambda\omega \exp[-\omega/\omega_{{\rm c}}]$, where $\lambda$ is the coupling strength and $\omega_{{\rm c}}$ is the cutoff frequency.
For the numerical calculation, we chose $2M$, the energy difference between the spin-$\uparrow$ and -$\downarrow$ states in the dot, as an energy unit. 
We distinguish parameters normalized by their units with an overbar (see note \cite{note:unit}).
Specific values of the normalized parameters are given in the figure captions.
As we are focusing on the spin transfer driven by the rotating magnetization, the dot is set in a steady state \cite{note6} at ${\bar t}=0$ to exclude any transient spin transfer caused by the dot--lead contact.
Under this initial condition, the net charge transfer $\langle\Delta n_{\uparrow}\rangle+\langle\Delta n_{\downarrow}\rangle$ is zero because the charge is conserved in the lead.
Nevertheless, a spin current $I_{{\rm spin}}$ is generated because equal amounts of spin-$\uparrow$ and spin-$\downarrow$ electrons are transferred in opposite directions, i.e., $\langle\Delta n_{\uparrow}\rangle=-\langle\Delta n_{\downarrow}\rangle$ because spin flips in the dot are driven by the rotating magnetization

\begin{figure}[t]
\centering
\includegraphics[keepaspectratio, scale=0.54,angle=0]{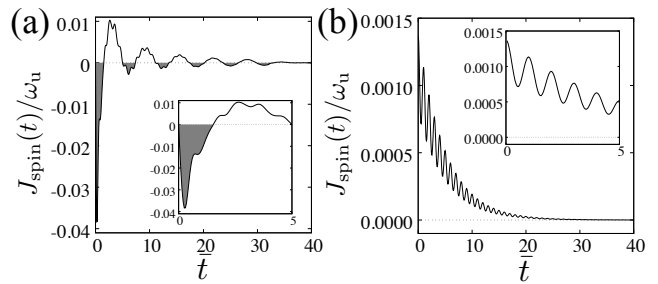}\label{fig2}
\vspace{-22pt}
\caption{Instantaneous spin current in a single interval under non-Markovian (a) and Markovian (b) dynamics; the insets are magnifications of the time interval $0\leq{\bar t}\leq5$. 
In each panel, $\phi$ is suddenly changed at ${\bar t}=0$ from $0$ to $\delta\phi$, then $\phi=\delta\phi$ is held fixed during the time interval. Before the sudden change at ${\bar t}=0$, the dot is in the steady state.
(a) The non-Markovian result with $J_{{\rm spin}}(0)=0$ at ${\bar t}=0$ and the frequent reversals of sign of $J_{{\rm spin}}({\bar t})$ marked as gray areas. The sign changes indicate backflow. (b) The Markovian result with $J_{{\rm spin}}(0)\not=0$ at ${\bar t}=0$ and $J_{{\rm spin}}({\bar t})>0$ for ${\bar t}>0$, indicating a monotonic transfer of spin. The parameters are set to ${\bar\epsilon}_{{\rm d}}=10$, ${\bar\mu}=10$, ${\bar\beta}=100$, $\lambda=0.01$, ${\bar\omega}_{c}=4$, $\theta=3\pi/4$, and $\delta\phi=\pi/10$. 
Dependence on the parameter choice is summarized in note\cite{note7}.
The time evolution is independent of $\phi$ because the system has rotational symmetry about the $z$-axis.}
\vspace{-15pt}
\end{figure}
Let us first examine the instantaneous spin current $J_{{\rm spin}}(t)$, Eq.~(\ref{eq:spinflow}).
We plot its time evolution under the non-Markovian analysis [Fig.~2(a)] as well as the corresponding Markovian analysis [Fig.~2(b)].
Both time evolutions are given for a single time interval for a step-like rotation of the magnetization (see the figure caption).
We set the interval to be larger than the relaxation time (specifically, ${\bar\tau}_r\sim20$ and $\delta{\bar t}=40$).
As the dot is initially in the steady state, the time evolution of $J_{{\rm spin}}(t)$ is driven by the sudden change of $\phi$ at ${\bar t}=0$.

Fig.~2(a) exhibits two different oscillations; the larger oscillation with the longer period reflects the back-and-forth transfer of spin between dot and lead caused by the non-Markovian dynamics arising from the dot--lead coupling, whereas the smaller oscillation with the shorter period reflects the periodic transition between the spin-$\uparrow$ and -$\downarrow$ states in the dot with Larmor frequency $2M/\hbar$.
In contrast, Fig.~2(b) only exhibits the Larmor precession.

Fig.~2(a) also shows that $J_{{\rm spin}}(t)$ under non-Markovian dynamics starts from zero at ${\bar t}=0$, which properly reflects the moment when the dynamics starts from the steady state.
The gray-colored region identifies negative spin current, $J_{{\rm spin}}(t)<0$, which we call spin backflow, where the spin current flows back from the lead.
In contrast, regarding the Markovian dynamics [Fig.~2(b)], we find that the spin starts flowing with a finite impetus at ${\bar t}=0$, always taking positive values during its time evolution, which indicates that the spin is always transferred from dot to lead without backflow.
Focusing on the initial short-time behavior, the direction of the instantaneous spin current in the non-Markovian dynamics ($J_{{\rm spin}}<0$) is opposite to that in the Markovian dynamics ($J_{{\rm spin}}>0$).
We examine the difference in detail in Supplementary Material.

\begin{figure}[t]
\centering
\includegraphics[keepaspectratio, scale=0.44, angle=0]{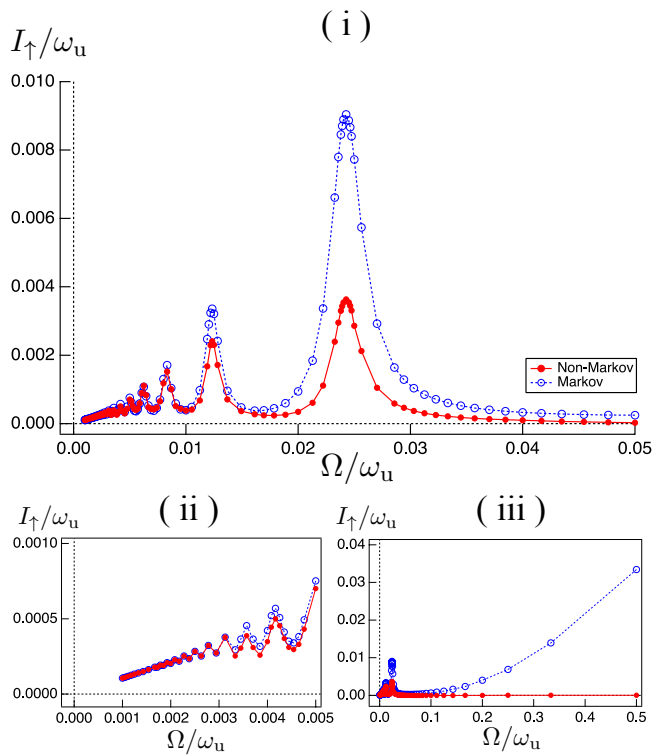}\label{fig1}
\vspace{-22pt}
\caption{Frequency dependences of the temporal average of spin current $I_{{\rm spin}}$. With fixed $\delta\phi=\pi/10$, the frequency is changed by changing $\delta t$. The red and blue dashed lines mark the non-Markovian and Markovian results, respectively.
Panel (i) presents frequency range $0\leq{\bar\Omega}\leq0.05$; panel (ii) presents a magnification of the range $0\leq{\bar\Omega}\leq0.005$; panel (iii) presents the dependence up to ${\bar\Omega}=0.5$.
Both results exhibit oscillations that depend on ${\bar\Omega}$ for ${\bar\Omega}\gtrsim0.002$ and is a consequence of Rabi oscillations in the dot. The results coincide in the linear regime (${\bar\Omega}\lesssim0.002$) whereas they deviate in the oscillating regime. The parameter values are the same as in Fig.~2.}
\vspace{-15pt}
\end{figure}
Let us now examine how the difference in the temporal behavior of $J_{{\rm spin}}(t)$ is reflected in the total spin current generation.
For the purpose, we evaluated the frequency dependence of the temporal average of spin current $I_{{\rm spin}}$, Eq.~(\ref{eq:spincurrent}), under non-Markovian and Markovian dynamics (Fig.~3).

For both dynamics, we find a common feature, i.e., the linear dependence on ${\bar\Omega}$ gradually changes to an oscillatory dependence for higher frequencies (around ${\bar\Omega}\gtrsim0.002$ in Fig.~3), which is explained by comparing the time interval $\delta{\bar t}$ and the relaxation time ${\bar\tau}_{r}$.
For lower frequencies, for which $\delta{\bar t}\gg{\bar\tau}_{r}$, the numerator of Eq.~(\ref{eq:spincurrent}) becomes constant because the instantaneous spin current $J_{{\rm spin}}(t)$ has already vanished at a certain ${\bar t}<\delta{\bar t}$ (see Fig.~2), which results in the linear dependence of $I_{\uparrow}$ on ${\bar\Omega}$.
As ${\bar\Omega}$ becomes larger and the time interval satisfies $\delta{\bar t}\lesssim{\bar\tau}_{r}$, the angle $\phi$ changes during relaxation.
In this situation, we have two extreme features; when $\delta{\bar t}$ is an integer multiple of the period of a spin flip $\hbar/2M$, we have resonance enhancement of the spin flip by changing $\phi$ to exhibit a maximum, whereas it is anti-resonantly suppressed to display a minimum when $\delta{\bar t}$ is a half-integer multiple of the period \cite{note8}.

\red{Comparing both analyses, we find a coincidence in the lower frequency (linear) regime (see panel (ii)), whereas they deviate over the higher frequency regime.
The coincidence is caused by the electron dynamics being well described with the Markovian approximation because, in the linear regime, the time interval $\delta t$ is sufficiently larger than the relaxation time as the long-time (Markovian) limit on the memory kernel is valid.
In contrast, in the higher frequency regime where $\delta t$ is small, the Markovian approximation breaks down, and the non-Markovian effect, specifically backflow, reduces the amount of $I_{{\rm spin}}$.
The deviation is quite significant in panel (iii); the Markovian analysis diverges with respect to $\Omega$, whereas the non-Markovian analysis is totally suppressed.
The divergence is unphysical as it is caused by the accumulation of the non-zero impetus of $J_{{\rm spin}}(t)$ just after the sudden change in $\phi$ under the Markovian analysis, which is an error caused by the Markovian approximation (see Fig.~2).}

\section{Discussions}
In the context of spin pumping in bulk systems, some researchers have studied the ``backflow (or backscatter) of the spin current" because of the {\it finite size} of the electron reservoir and the slow modulation of the system to follow the precession sufficiently \cite{tserkovnyakprb02,chen15}.
They have argued that, when the pumped angular momentum does not quickly dissipate to the lead, a nonvanishing spin accumulation may build up in the lead.
For a sufficiently slow precession, the spin imbalance through spin accumulation may flow back into the ferromagnet, canceling the generated spin current as the system is always in a steady state.
The behavior of this backflow in spin current is different from the spin backflow studied in this work in regard to two points: (i) the latter occurs even for an ideal reservoir in which the pumped spin is absorbed entirely, whereas the former is caused by the accumulation of spin angular momentum in the finite reservoir, and (ii) the latter becomes significant for rapid precession, whereas the former requires a sufficiently slow precession. 
Therefore, the spin backflow studied in this paper is a completely independent concept from the conventional backflow of spin current.
When one considers a non-ideal reservoir of finite size and a moderately rapid precession, both backflow processes may coexist.
A study of the situation is left for a future investigation.

Although we have focused on the spin backflow in this work, the concept of backflow itself is a universal feature of quantum transport in non-Markovian dynamics.
Indeed, some researchers have studied the backflow of information \cite{breuer16} and energy \cite{guarnieri16} in undriven systems.
Because this is the first study of backflow in a driven system, we conjecture that our main result, the reduction of the pumped quantity because of backflow, holds for a wide range of driven systems.
We shall discuss the universality of our results elsewhere.

The steplike rotation reduces to a continuous rotation in a limit $\delta t\to0$, $\delta\phi\to0$ with $T={\rm constant}$.
With a non-zero Markovian flow at $t=0$, the limit leads to a divergence of the spin current under the Markovian approximation (Fig.~3).
To avoid this divergence, we need to include the non-Markovian effect.

\section{Conclusions}
Focusing on spin backflow, we have examined the role of the non-Markovian effect on the spin pumping under a precessing magnetization.
In evaluating the frequency dependence of the pumped spin current, we compared the results obtained from our non-Markovian analysis with those under a corresponding Markovian analysis.
Our numerical result shows that spin backflow does not contribute to the net amount of spin current in the low-frequency regime where $\delta t\gtrsim\tau_{r}$, whereas it significantly reduces the spin current in the high-frequency regime where $\delta t\lesssim\tau_{r}$.
This provides a physically reasonable description of spin pumping over all frequencies, which a conventional Markovian approximation is unable to achieve.

\section*{Acknowledgments}
This work was supported by the Grant-in-Aid for Challenging Exploratory Research (No. 16K13853) and partially supported by the Grant-in-Aid for Scientific Research on Innovative Areas Science of Hybrid Quantum Systems (No. 18H04290).


\begin{thebibliography}{99}
\bibitem{tatara_note}
This understanding is consistent with the fact that the spin pumping effect is induced by a non-adiabatic component of the effective gauge field\cite{tatara19}.
The linear regime is sometimes referred to as the "adiabatic response"\cite{avron12}.
\bibitem{note}
A typical single molecule magnet studied in the molecular spintronics is octanuclear iron(III) oxo-hydroxo, $[{\rm Fe_{8}O_{2}(OH)_{12}(tacn)_{6}}]^{8+}$ (in short, ${\rm Fe}_{8}$). Following Ref.~\cite{misiorny07}, the coupling strength between its lowest unoccupied molecular orbital (LUMO) and an external electron lead is $\Gamma\approx0.0015$~meV. Therefore, the rate of relaxation by tunneling is estimated to be $\gamma_r\approx0.36$~GHz. Assuming the precession of its magnetic core is excited by microwaves, a typical frequency of which is several tens of gigahertz, the precession frequency $\nu$ is more than ten times the relaxation rate $\gamma_r$.
\bibitem{note4}
For $\lambda_{\sigma}=0$, both the operator $\rho^{(\lambda_{\sigma})}(t)$ and the equation coincide with the reduced density matrix of the dot $\rho^{(0)}(t)\equiv{\rm Tr}_{{\rm l}}[W(t)]$, where ${\rm Tr}_{{\rm l}}$ is the trace taken over the lead and $W(t)$ is the density matrix of the total system, and to the master equation for $\rho^{(0)}(t)$ of the time convolutionless type.
\bibitem{note:unit}
We introduce a unit energy $\epsilon_{{\rm u}}\equiv2M$, a unit angular frequency $\omega_{{\rm u}}\equiv2M/\hbar$, and a unit time $t_{{\rm u}}\equiv2\pi/\omega_{{\rm u}}$, and define the normalized quantities of energy, angular frequency, inverse temperature, and time 
${\bar\omega}\equiv\omega/\omega_{{\rm u}}$, 
${\bar\omega}\equiv\omega/\omega_{{\rm u}}$, 
${\bar\epsilon}\equiv\epsilon/\epsilon_{{\rm u}}$, 
${\bar\beta}\equiv\beta/\beta_{{\rm u}}$, and 
${\bar t}\equiv t/t_{{\rm u}}$, respectively.
\red{Experimentally, the unit energy is $2M\approx0.025~{\rm meV}$ for ${\rm Fe}_{8}$ single molecule magnet\cite{misiorny07}.
With this unit energy, the unit angular frequency and the unit time are $2M/\hbar\approx38\times10^{9}~{\rm s}^{-1}$ and $2\pi/\omega_{{\rm u}}\approx0.17~{\rm ns}$, respectively.}
\bibitem{note5}
The parameters satisfy conditions $\epsilon_{{\rm d}}-M<\mu<\epsilon_{{\rm d}}+M$ and $\beta^{-1}\lesssim2M$, which are essential for spin pumping because, if they are not satisfied, either electrons do not transfer to the dot (as $\epsilon_{{\rm d}}-M>\mu$) or spin-$\uparrow$ and spin-$\downarrow$ electrons of equal amounts flow onto the dot (for $\mu>\epsilon_{{\rm d}}+M$) \cite{hashimoto17}.
\bibitem{note6}
The steady state $\rho^{{\rm st}}$, satisfying $\xi^{(0)}(t\to\infty)\rho^{{\rm st}}=0$, is analytically obtained using a graphical method discussed in Ref.~\cite{Haken}. An analytical expression for the minimum model is provided in Appendix C of our previous paper Ref.~\cite{hashimoto17}.
\bibitem{note7}
\red{(A) The amount of spin backflow monotonically increases as the cutoff frequency $\omega_c$ increases and, it saturates towards a certain value in the limit $\omega/\omega_c\to0$. This is because the dot--lead coupling represented by $v((\epsilon_{{\rm d}}\pm M)/\hbar)$ increases towards the Ohmic (linear) function $\lambda(\epsilon_{{\rm d}}\pm M)/\hbar$ in the limit.
(B) The spin current decreases as the lead temperature increases because the numeric difference between the transferred electrons with spin-$\uparrow$ and spin-$\downarrow$ approaches zero, for the reason that the two electron populations in the lead during the interaction windows of the spin-$\uparrow$ and -$\downarrow$ electrons move closer.
(C) The spin polarization of the current exhibits a $\theta$ dependence in that for $0<\theta<\pi/2$ the spin polarization is antiparallel to the $z$-axis, whereas for $\pi/2<\theta<\pi$ the spin polarization is parallel to the $z$-axis for the non-Markovian analysis; see \cite{hashimoto17} for a Markovian analysis.}
\bibitem{note8}
As the maxima and minima are determined by the timing of the change of $\phi$ (or $\delta t$) and the Rabi period in the dot, $\hbar/2M$, they appear at the same frequencies in both the non-Markovian and Markovian results. For the Markovian analysis, we discussed in detail the oscillatory behavior in Ref.~\cite{hashimoto17}, Sec. 2.2. The discussion holds qualitatively for the non-Markovian analysis.
\end{thebibliography}
\end{document}